\documentclass{elsart}

\usepackage{natbib}

\usepackage{psfig}
\usepackage{wrapfig}

\usepackage{amssymb}

\def\kms{km~s$^{-1}$}
\def\la{\ifmmode\stackrel{<}{_{\sim}}\else$\stackrel{<}{_{\sim}}$\fi} 
\def\ga{\ifmmode\stackrel{>}{_{\sim}}\else$\stackrel{>}{_{\sim}}$\fi}

\begin{document}

\begin{frontmatter}



\title{BOW SHOCKS AROUND PULSARS AND NEUTRON STARS}


\author{Bryan M. Gaensler}

\address{Harvard-Smithsonian Center
for Astrophysics, 60 Garden Street MS-6, Cambridge, MA 02138, USA}

\begin{abstract}
Pulsar wind nebulae are now well established as important probes both of
neutron stars' relativistic winds and of the surrounding interstellar
medium.  Amongst this diverse group of objects, pulsar bow shocks have
long been regarded as an oddity, only seen around a handful of rapidly
moving neutron stars.  However, recent efforts at optical, radio and
X-ray wavelengths have identified many new pulsar bow shocks, and these
results have consequently motivated renewed theoretical efforts to model
these systems.  Here I review the new results and ideas which have
emerged on these spectacular systems, and explain how bow shocks
and ``Crab-like'' nebulae now form a consistent picture within our
understanding of pulsar winds.
\end{abstract}

\begin{keyword}

ISM: general \sep
stars: neutron \sep
stars: winds, outflows



\end{keyword}

\end{frontmatter}

% main text
\section{Introduction}
\label{}

Pulsars have high space velocities, typically in the range 150--1500~\kms\
\citep*{acc02}. Since such speeds are well in excess of typical sound
speeds for the cold ($\sim1$~\kms) or warm components ($\sim10$~\kms)
of the interstellar medium (ISM), we expect that many pulsars are moving
supersonically, and that they should thus drive bow shocks through
ambient gas.

Such phenomena are, of course, not unique: astrophysical bow shocks are
seen around proto-stars, clumps of supernova ejecta, runaway OB stars,
and in the solar wind \citep[e.g.,][]{kva+97,vm88b,sd95b,zan99}. However,
several factors make pulsar bow shocks particularly appealing to study.
First, for pulsars the mechanical luminosity of the wind driving the
bow shock can be reasonably estimated through the pulsar's ``spin-down
luminosity'', given by $\dot{E} \equiv 4 \pi^2 I \dot{P}/{P^3}$ (where
$I \equiv 10^{45}$~g~cm$^2$ is the  neutron star's assumed moment of
inertia, $P$ is the star's spin period, and $\dot{P}$ is the period
derivative). Spin-down luminosities for observed pulsars are typically
in the range $10^{32}-10^{39}$~ergs~s$^{-1}$. Second, for many pulsars,
precision timing and interferometric measurements allow accurate
determinations of the star's position and proper motion. Finally,
compared to many other Galactic populations, the distances to pulsars
are reasonably well-established --- from parallax for about 20 sources
\citep[e.g.,][]{ccv+04}, and from dispersion of the pulsed radio signals
for over 1500 others \citep{cl02}.  All these factors  mean that many
of the unknowns generally associated with bow shocks are not an issue
for pulsars; in principle, the only parameters to be determined are
the density and structure of the ISM, plus the inclination angle of
the system.

Furthermore, most pulsars are $10^6-10^9$ years old, their high space
velocities having carried them far from their birth places.  This means
that pulsars are more or less randomly located within the Galactic disk,
and that their bow shocks can thus act as unbiased probes of the ISM.
Bow shocks also potentially provide an opportunity to study the winds of
``typical'' pulsars, rather than extreme cases of youth and energy as
typified by the Crab Nebula.

Despite their potential as useful probes of pulsars and their
surroundings, very few pulsar bow shocks were known until recently.
Because of several recent discoveries and accompanying theoretical
modelling, there has been renewed interest in these systems in the last
few years. Here I summarise recent work, and the resulting improvements
to our understanding. For earlier reviews on bow shocks see \cite{cor96}
and \cite{cc02}; for more general discussion of pulsar wind nebulae,
the reader is referred to reviews by \cite{gae04b} and \cite{krh04}.

\section{Theoretical Expectations}
\label{sec_theory}

Several authors have considered the theory of bow shocks in detail, both
under general circumstances \citep{cbw96,wil96,wil00} and specifically
in the case of pulsars \citep{arc92,buc02a,buc02b,vag+03}. Some pertinent
points from these studies are as follows.

\begin{figure}[htb]
%\centerline{\psfig{file=fig1.eps,width=\textwidth,clip=}}
\vspace{6cm}
\caption{Hydrodynamic simulation of a pulsar bow shock, adapted
from \cite{buc02a}. The greyscale indicates density; the pulsar
is moving from right to left with a Mach number $\mathcal{M} \sim 13$. 
Various regions and interfaces
discussed in the text are indicated.}
\label{fig_bucc}
\end{figure}

{\bf Scaling Parameter:}  One parameter sets the overall scale of a
bow shock --- the ``stand-off distance'', $r_w$. This is the separation
between the pulsar position and the apex of the bow shock, and is set
by pressure balance, 
\begin{equation}
\frac{\dot{E}}{4\pi r_w^2 c} = \rho V^2,
\label{eqn_balance}
\end{equation}
where $\rho$ is the ambient mass density and $V$ is the pulsar's space
velocity relative to its surroundings.

{\bf Shape:} \cite{wil96} has provided an elegant analytic solution to the shape
of a bow shock, 
\begin{equation}
r(\theta) = r_w \csc \theta \sqrt{3(1-\theta \cot \theta)},
\label{eqn_wilkin}
\end{equation}
where $r(\theta)$ is the radius of the bow shock at a polar
angle $\theta$. This solution is for an idealised thin-layer shock, but
is also expected to be a reasonable approximation to pulsar bow shocks
in regions near the apex \citep{buc02a,vag+03}.

{\bf Overall Structure:} Because of the long cooling time scales,
bow shocks are not thin shells, but show a characteristic double-shock
structure, as shown in Figure~\ref{fig_bucc}. The morphology consists of a
forward shock which heats the ISM, a termination shock which decelerates
the pulsar wind, and a contact discontinuity dividing shocked ISM from
shocked pulsar wind material.

{\bf Termination Shock:} The pulsar wind experiences a significant
difference in pressure between regions ahead of and behind the
pulsar's motion. The termination shock is therefore not of uniform
radius around the pulsar, but rather has a bullet-shaped morphology,
the head of the bullet aligned with the pulsar's direction of motion
as can be seen in Figure~\ref{fig_bucc}.
For low Mach numbers, $\mathcal{M} \sim 1$,
the ratio of termination shock radii between the directions immediately
behind and directly ahead of the pulsar is roughly proportional to 
$\mathcal{M}$
\citep{buc02a,vag+03}. However, in the limit of high Mach number, this
ratio tends to an asymptotic limit of about 5 \citep{gvc+04}.

\section{Observations: Forward Shock}
\label{sec_fwd}

The forward shock should produce observable H$\alpha$ emission,
resulting from collisional excitation and charge exchange of neutrals
in ambient gas. Indeed some beautiful bow shocks have been observed in the
H$\alpha$ line, several of which have been discovered just in the last few
years \citep{vk01,jsg02,gjs02}. Detection of such a bow shock allows
one to estimate measure $r_w$, although two cautions must be issued.
First, $r_w$ formally corresponds to the distance from the pulsar to
the termination shock, not to the forward shock \citep[e.g.,][]{buc02a}.
So the observed pulsar/shock separation needs to be scaled by some factor,
which simulations suggest has a value $\sim0.4-0.6$ \citep{buc02a,vag+03}.
Second, the measured separation needs to be corrected for the (usually
unknown) inclination.  This correction is not a simple trigonometric
factor, because projection of the three-dimensional bow-shock surface
onto the sky can be larger than the true separation \citep{gjs02}.
Correction for inclination is non-trivial, and quantities inferred from
the observed bow-shock separation should thus be considered approximate.

With these caveats in mind, pressure balance can be applied to
a system in which $\dot{E}$ and $V$ are known to yield $\rho$ via
Equation~(\ref{eqn_balance}). This approach suggests ambient densities
$\sim0.1$~cm$^{-3}$, as expected for warm neutral gas in the ISM
\citep{cc02,gjs02}. In cases where $V$ is not known, one can write
$\rho V^2 = \gamma \mathcal{M}^2 \mathcal{P}$, where $\gamma = 5/3$
is the adiabatic coefficient of the ISM and $\mathcal{P}$ is the ISM
pressure. Adoption of a typical value for $\mathcal{P}$ allows one to
directly estimate $\mathcal{M}$.

In the case of an isotropic pulsar wind propagating through a homogeneous
ISM, Equation~(\ref{eqn_wilkin}) should describe the H$\alpha$
brightness profile well. For PSR~J0437--4715 
this indeed seems to be the case, allowing one to solve
for the inclination angle of the system \citep{mrf99}. However, 
in several other
cases there are significant deviations from this idealised solution,
in the form of abrupt kinks and bulges in the bow-shock profile, or
a rotational offset between the symmetry axis of the bow shock and the
velocity vector of the pulsar (e.g., PSR~B0740--28, \citeauthor{jsg02}
\citeyear{jsg02}; PSR~J2124--3358, \citeauthor{gjs02} \citeyear{gjs02}).
These shapes imply some combination of anisotropies in the pulsar
wind (as are known to exist around young pulsars; see \citeauthor{gae04b}
\citeyear{gae04b}), density
variations in the ISM, or significant motion of ambient gas with respect
to the local restframe. Multi-epoch observations, showing changes in
the bow-shock structure with time, are beginning to explore these
possibilities \citep{cc04}.

\section{Observations: Termination Shock}
\label{sec_term}

\subsection{The Mouse}

Just as is the case for younger systems like the Crab Nebula, we
expect that the particles in the pulsar wind will be accelerated at
the termination shock, generating radio and X-ray synchrotron emission.
Indeed cometary radio and X-ray emission aligned with the direction of
motion is seen around several high velocity pulsars \citep{cc02}.

\begin{figure}[htb]
%\centerline{\psfig{file=fig_mouse_xfig.eps,width=0.9\textwidth}}
\centerline{\psfig{file=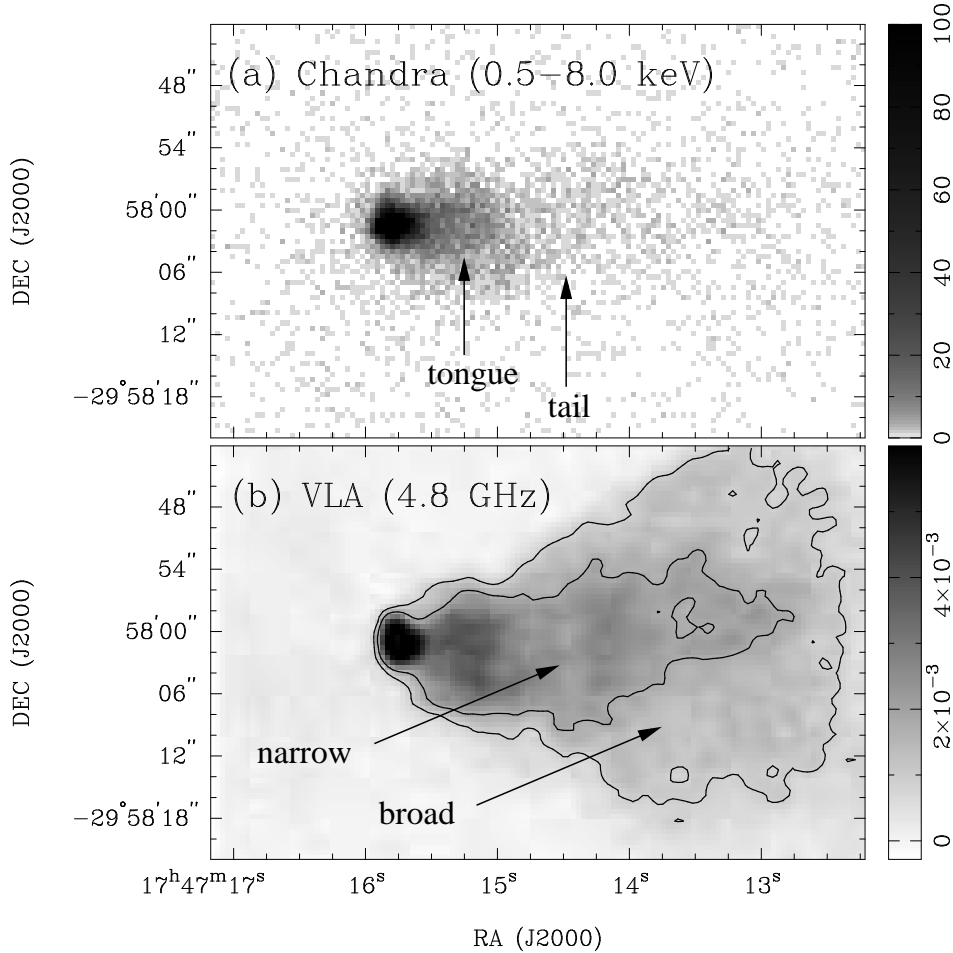,width=0.9\textwidth}}
\caption{X-ray (upper) and radio (lower) images of
the bow shock G359.23--0.82 (a.k.a.\ ``the Mouse'') powered
by PSR~J1747--2958 \citep{gvc+04}. In the X-ray image, the ``tongue''
and ``tail'' regions are indicated. In the radio image, the
broad and narrow components of the tail are shown.}
\label{fig_mouse}
\end{figure}

Figure~\ref{fig_mouse} shows the results of a recent detailed study of
``the Mouse'', an elongated radio and X-ray nebula coincident
with the energetic pulsar J1747--2958 \citep{cmgl02,gvc+04}.  The Mouse
is very luminous, and thus represents an ideal opportunity to test the
theories developed for bow-shock termination shocks.

If the X-ray and radio emission from the Mouse comes from the termination
shock region, we expect it to be sharply bounded at its apex by
the contact discontinuity. The extent of emission thus allows us to
estimate the stand-off distance for this system --- for a distance of
5~kpc we find $r_w \approx 0.02$~pc.  For ISM pressures $\mathcal{P}/k
= 2400P_0$~K~cm$^{-3}$ (with typical observed values in the range
$0.5 \la P_0 \la 5$), this implies a high Mach number, $\mathcal{M}
\sim 60P_0^{-1/2}$, and thus a likely space velocity $V \approx
600P_0^{-1/2}$~\kms\ if propagating through the warm component of the ISM.

The X-ray and radio morphologies shown in Figure~\ref{fig_mouse} both
indicate the presence of an elongated bright ``tongue'' of emission
behind the pulsar, about 0.25~pc in length.  An interpretation of this
feature is suggested by comparison with the hydrodynamic simulation
shown in Figure~\ref{fig_bucc}, where it can be seen that the tongue
has a similar appearance to the surface of the bullet-shaped termination
shock. If we make this identification, then the tongue in the Mouse is
directly analogous to the inner toroidal ring seen for the Crab Nebula
and other related systems \citep[e.g.,][]{nr04}, demarcating the point
where the pulsar wind reacts to its surroundings and where wind particles
are accelerated up to synchrotron-emitting energies.

It is important to note that hydrodynamic simulations such as that shown
in Figure~\ref{fig_bucc} do not explicitly include the effects of magnetic
fields or relativistic flows, so an exact correspondence between the data
and the model is not expected.  In particular, two important discrepancies
can be identified: the ratio of backward and forward termination shocks
is observed to be $\sim10$, rather than the value of 5 predicted for $\mathcal{M}
\gg 1$; and the sheath does not show any limb-brightening as would be
expected given its geometry. The first of these points can potentially
be explained by the effect of ions in the wind, whose gyrations in
the magnetic field spread the electron shock over a large distance 
\citep{ga94}. This effect should operate in
the backward flow from the pulsar, but not in the forward
direction due to the high degree of pressure confinement.  The lack
of limb brightening
can be accounted for by Doppler boosting in the post-shock flow,
as is consistently observed in the winds of younger, lower velocity
pulsars \citep{nr04}.  While confirmation of these effects will require
a fully relativistic magnetohydrodynamic treatment, we emphasise that
both the effects claimed here have been seen in the wind nebulae around
``Crab-like'' pulsars. The need to invoke similar phenomena here suggests
a high degree of commonality between the wind properties of the youngest
pulsars and more typical members of the population.

West of the tongue, a fading tail of emission is seen, which
must come from
material significantly downstream of the termination shock. 
This emission is analogous to the
main body of the nebula in Crab-like systems.  The tail
of emission behind the Mouse can be delineated into two components, as
indicated in Figure~\ref{fig_mouse}.  The first component is bright,
reasonably narrow, and is seen in both radio and X-rays; the second
component is broader, fainter, and is seen in radio only. These structures
can be qualitatively understood by again considering the hydrodynamic
simulation of Figure~\ref{fig_bucc}, in which it can be seen that there
are two extremes in the flow. Ahead of the pulsar, the wind shocks at a
small distance from the pulsar at the apex, then turns around and flows
around the edges of the shocked region. In contrast, behind the pulsar
the distance to the shock is considerably larger; the post-shock flow
then remains in a narrow collimated region lying along the symmetry
axis. If equipartition holds, so that $\dot{E}/4\pi r_w^2
c \propto B^2$, where $B$ is the post-shock nebular magnetic field,
then the shocked flow ahead of the pulsar should have a magnetic field
$\sim10$ times larger than that behind the pulsar. We thus expect
a broad flow in which the X-ray emission suffers severe synchrotron
losses, enveloping a narrow flow in which the X-rays are yet to cool.
This is in accordance with observations, suggesting that in contrast to
the Crab Nebula, the pulsar's motion structures the post-shock emitting
regions into two distinct zones.

\subsection{X-ray Trails Behind Other Pulsars}

Now that we have a picture of what might be occurring for the
Mouse, it is insightful to consider other bow shocks which produce radio
and X-ray emission, to see if similar features are observed.

Some pulsars, such as PSRs~B1757--24 and B1957+20 \citep{kggl01,sgk+03},
show short ($\sim0.1-0.5$~pc) X-ray trails extending opposite the pulsar's
direction of motion.  One interpretation is that these features are
synchrotron ``wakes'', i.e., particles left behind by the pulsar as it
moves through the ISM.  However, the time taken for a pulsar to traverse
the extent of one of these trails is $\sim1000$~yr, implying abnormally
low nebular magnetic field strengths if synchrotron cooling at X-ray
energies is to be avoided during this period.  It has therefore been
suggested that a rapid backflow or nozzle operates downstream, which
rapidly advects emitting particles and produces the trail structures
observed \citep[e.g.,][]{wg98b,kggl01}.

However, the Mouse suggests an alternative interpretation.  We expect
two components to the downstream synchrotron emission from a pulsar
bow shock. Close to the pulsar, we should see a ``tongue'' of emission
around the termination shock. We expect this region to show a constant
X-ray spectrum across its extent, to have a comparable extent in radio
and X-rays, and to abruptly terminate at a distance $\sim10r_w$ behind
the pulsar. Further from the pulsar, the ``tail'' component representing
post-shock material should show an X-ray spectrum which softens with
distance from the pulsar, should have a larger extent in radio than in
X-rays, and should gradually fade into the background over an extent
$\gg10r_w$.

Armed with the realisation that bow shocks should show both these
components, it seems that the short trails seen around PSRs~B1957+20
and B1757--24 correspond well to the tongue component of the Mouse.
\cite{gvc+04} and \cite{gva04b} have thus both argued that the name
``trail'' is a misnomer --- these features are simply the surfaces of
the termination shock around these pulsars, the longer tail
of emission downstream being too faint to see.

%\begin{wrapfigure}{l}{0.8\textwidth}
\begin{figure}[t!]
%\centerline{\psfig{file=fig_ic443_xfig.eps,width=0.8\textwidth}}
\centerline{\psfig{file=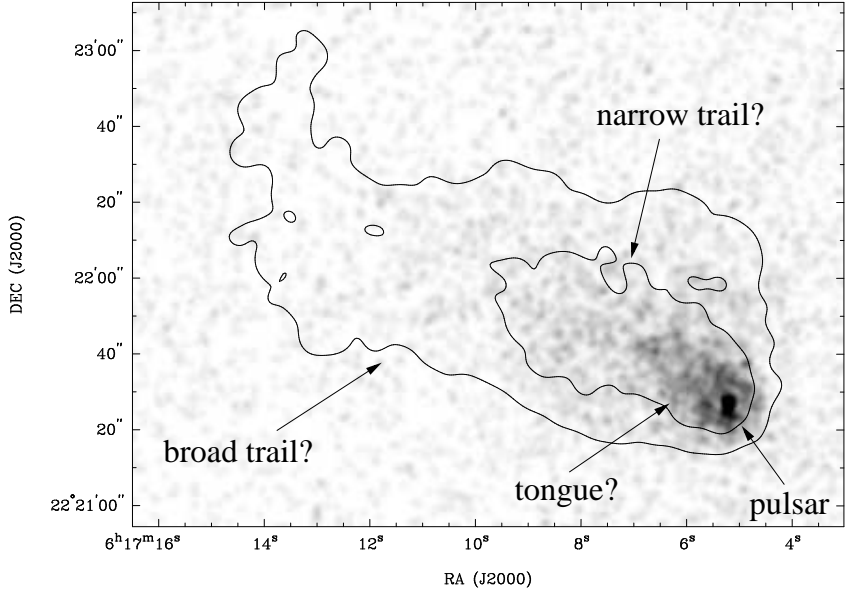,width=0.8\textwidth}}
\caption{X-ray 
and radio observations of the X-ray bow shock around
CXOU~J061705.3+222127 in the SNR~IC~443. 
The image shows {\em Chandra}\ data 
smoothed with a $2''$ gaussian,
while contours denote 8.4-GHz VLA data at a resolution of $7''$.
Proposed features are indicated.}
\label{fig_ic443}
\end{figure}
%\end{wrapfigure}

On the other hand, a closer examination of other pulsar bow shocks
suggests that they too potentially show multiple components in X-rays,
corresponding both to the termination shock and to the tail of emission
behind it. Possible examples of such systems are PSR~B1853+01 in the
supernova remnant (SNR) W44 \citep{pks02}, PSR~B1951+32 in SNR~CTB~80
\citep{mle+04}, and the neutron star CXOU~J061705.3+222127 in SNR~IC~443
\citep{ocw+01}.  The latter source is shown in Figure~\ref{fig_ic443},
where the identification of these possible structures is indicated. It
is interesting to note that this source is moving through the shocked
gas in the interior of an evolved SNR. In this case, we expect a
Mach number $\mathcal{M} \sim 3$ \citep{vag+03}, much lower than for a
pulsar moving through the ISM.  For this low Mach number, we expect the
``tongue'' region to only be about half as long (in units of $r_w$)
than for a high Mach number system like the Mouse. Indeed the data in
Figure~\ref{fig_ic443} suggest an extent for this feature $\sim5r_w$,
compared to $\sim10r_w$ for the Mouse.  Deeper {\em Chandra}\ observations
of both CXOU~J061705.3+222127 and PSR~B1853+01 are scheduled to take place
in 2005, which should allow a better investigation of these possibilities.

\section{Putting It All Together}

In \S\ref{sec_theory}, it was noted that a double-shock structure is
expected around a supersonic pulsar. While we clearly see either forward
or termination shocks around various neutron stars, 
for just one system, PSR~B1957+20, have both the termination
shock (in X-rays) and forward shock (in H$\alpha$) been identified in
the same source \citep{sgk+03}.  A collection of such sources can act
as a detailed probe of the hydrodynamics of the bow-shock interaction,
so expansion of this sample is obviously highly desirable.

Most of the pulsar bow shocks seen in radio/X-rays have high values of
$\dot{E}$, meaning that they are reasonably rare and hence distant.
The high extinction to these systems makes it difficult to identify
any H$\alpha$ emission around them. On the other hand, H$\alpha$ bow
shocks tend to be nearby, so are worthwhile targets for X-ray imaging,
even if $\dot{E}$ is low.  But of the five other known H$\alpha$
bow shocks besides that around PSR~B1957+20, two (PSR~J0437--4715,
\citeauthor{zps+02} \citeyear{zps+02}; RX~J1856.5--3754\footnote{This
source is one of several nearby
isolated neutron stars seen in thermal X-rays, and does not pulse.
We assume that this
source is a typical pulsar but beaming away from us. To the best of
our knowledge, the lack of extended X-ray emission has not been stated
in the literature, but is immediately apparent from inspection of deep
archival {\em Chandra}\ observations of this field.}) show no extended
X-ray emission, one (PSR~B2224+65; \citeauthor{wcc+03} \citeyear{wcc+03})
shows an X-ray filament completely misaligned with the pulsar's direction
of motion, and the remaining two (PSRs~B0740--28 and J2124--3358)
are yet to be observed at high angular resolution by {\em Chandra}.
Radio searches to some of these sources have also not been successful
at detecting extended emission \citep{gsf+00,cc02}. The lack of emission
from the termination shock in systems for which the wind is known to be
strongly confined is puzzling.

\section{Conclusions}

A consistent understanding is now beginning to emerge of the various
multi-wavelength structures seen around high velocity pulsars.
As expected, we see optical emission from the shocked ISM in several bow
shocks,  from which we can infer the density of the ambient medium or the
Mach number of the system.  The morphologies of these shocks also allow
us to probe anisotropies in the ISM and in the pulsar wind.  

The X-ray
and radio emission produced at the termination shock in these systems
permit a study of the relativistic wind which drives the bow shock.
Just as in the Crab Nebula, high-resolution X-ray imaging allows us
to identify both the surface of the termination shock and emission
from the post-shock flow. As in the Crab, the effects of an ion-loaded
wind and of Doppler beaming are required to explain the observations,
suggesting that the wind composition and behaviour is similar for bow
shocks and for Crab-like systems.
However, a distinct difference from the Crab
is the presence of two flow zones downstream, corresponding to highly-
and weakly-magnetized material, emerging from the front and back of the
termination shock, respectively.

In the near future, we expect that deeper X-ray and optical
images will increase the sample and show new types of structure
\cite[e.g.,][]{cbd+03}, and that relativistic magnetohydrodynamical
simulations can begin to provide a solid theoretical underpinning for
the observed morphologies. The time domain is now also beginning to play
a role, in that we can watch bow shocks evolve as they trace the cascade
of density inhomogeneities in the ISM \citep{cc04}.  Clearly bow shocks
are now beginning to fulfil  their long-standing promise to act as unique
probes of pulsars and their surroundings.

\section*{Acknowledgments}

I thank all my collaborators with whom I've studied pulsar bow
shocks over the last few years. This work has been supported by NASA
through SAO grant GO2-3075X and LTSA grant NAG5-13032.

%\bibliographystyle{elsart-harv}
%\bibliography{journals,modrefs,psrrefs,crossrefs}

\end{document}